\documentclass[article,reprint,superscriptaddress,longbibliography]{revtex4-1}

\usepackage{graphicx}
\usepackage{dcolumn}
\usepackage{bm}
\usepackage{hyperref} 
\usepackage{breakurl}
\usepackage{multirow}
\usepackage{enumitem}
\usepackage{color}
\usepackage[pagewise]{lineno}
\setcitestyle{super} 

\begin{document}

\title{Ta$_{2}$Pd$_{3}$Te$_{5}$ topological thermometer}

\author{Yupeng Li}
      \thanks{Equal contributions}
      \affiliation{Beijing National Laboratory for Condensed Matter Physics, Institute of Physics, Chinese Academy of Sciences, Beijing 100190, China}

\author{Anqi Wang}
      \thanks{Equal contributions}
      \affiliation{Beijing National Laboratory for Condensed Matter Physics, Institute of Physics, Chinese Academy of Sciences, Beijing 100190, China}
      \affiliation{School of Physical Sciences, University of Chinese Academy of Sciences, Beijing 100049, China}

\author{Senyang Pan}
      \affiliation{Anhui Province Key Laboratory of Condensed Matter Physics at Extreme Conditions,
      High Magnetic Field Laboratory of the Chinese Academy of Sciences, Hefei 230031, Anhui, China}

\author{Dayu Yan}
      \affiliation{Beijing National Laboratory for Condensed Matter Physics, Institute of Physics, Chinese Academy of Sciences, Beijing 100190, China}

\author{Guang Yang}
      \affiliation{Beijing National Laboratory for Condensed Matter Physics, Institute of Physics, Chinese Academy of Sciences, Beijing 100190, China}

\author{Xingchen Guo}
      \affiliation{Beijing National Laboratory for Condensed Matter Physics, Institute of Physics, Chinese Academy of Sciences, Beijing 100190, China}
      \affiliation{School of Physical Sciences, University of Chinese Academy of Sciences, Beijing 100049, China}

\author{Yu Hong}
      \affiliation{Beijing National Laboratory for Condensed Matter Physics, Institute of Physics, Chinese Academy of Sciences, Beijing 100190, China}
      \affiliation{School of Physical Sciences, University of Chinese Academy of Sciences, Beijing 100049, China}

\author{Guangtong Liu}
      \affiliation{Beijing National Laboratory for Condensed Matter Physics, Institute of Physics, Chinese Academy of Sciences, Beijing 100190, China}
      \affiliation{Songshan Lake Materials Laboratory, Dongguan 523808, China}

\author{Fanming Qu}
      \affiliation{Beijing National Laboratory for Condensed Matter Physics, Institute of Physics, Chinese Academy of Sciences, Beijing 100190, China}
      \affiliation{School of Physical Sciences, University of Chinese Academy of Sciences, Beijing 100049, China}
      \affiliation{Songshan Lake Materials Laboratory, Dongguan 523808, China}

\author{Zhijun Wang}
      \affiliation{Beijing National Laboratory for Condensed Matter Physics, Institute of Physics, Chinese Academy of Sciences, Beijing 100190, China}
      \affiliation{School of Physical Sciences, University of Chinese Academy of Sciences, Beijing 100049, China}

\author{Tian Qian}
      \affiliation{Beijing National Laboratory for Condensed Matter Physics, Institute of Physics, Chinese Academy of Sciences, Beijing 100190, China}
      \affiliation{Songshan Lake Materials Laboratory, Dongguan 523808, China}

\author{Jinglei Zhang}
      \email{zhangjinglei@hmfl.ac.cn}
      \affiliation{Anhui Province Key Laboratory of Condensed Matter Physics at Extreme Conditions,
      High Magnetic Field Laboratory of the Chinese Academy of Sciences, Hefei 230031, Anhui, China}

\author{Youguo Shi}
      \email{ygshi@iphy.ac.cn}
      \affiliation{Beijing National Laboratory for Condensed Matter Physics, Institute of Physics, Chinese Academy of Sciences, Beijing 100190, China}
      \affiliation{Songshan Lake Materials Laboratory, Dongguan 523808, China}

\author{Li Lu}
      \email{lilu@iphy.ac.cn}
      \affiliation{Beijing National Laboratory for Condensed Matter Physics, Institute of Physics, Chinese Academy of Sciences, Beijing 100190, China}
      \affiliation{School of Physical Sciences, University of Chinese Academy of Sciences, Beijing 100049, China}
      \affiliation{Songshan Lake Materials Laboratory, Dongguan 523808, China}

\author{Jie Shen}
      \email{shenjie@iphy.ac.cn}
      \affiliation{Beijing National Laboratory for Condensed Matter Physics, Institute of Physics, Chinese Academy of Sciences, Beijing 100190, China}
      \affiliation{Songshan Lake Materials Laboratory, Dongguan 523808, China}

\date{\today}

\begin{abstract}
In recent decades, there has been a persistent pursuit of applications for surface/edge states in topological systems, driven by their dissipationless transport effects. However, there have been limited tangible breakthroughs in this field.
This work demonstrates the remarkable properties of the topological insulator Ta$_{2}$Pd$_{3}$Te$_{5}$, as a thermometer. This material exhibits a power-law correlation in temperature-dependent resistance at low temperatures, stemming from its Luttinger liquid behavior of edge states, while exhibiting semiconductor behavior at high temperatures.
The power-law behavior effectively addresses the issue of infinite resistance in semiconductor thermometers at ultra-low temperatures, thereby playing a crucial role in enabling efficient thermometry in refrigerators supporting millikelvin temperatures or below.
By employing chemical doping, adjusting thickness, and controlling gate voltage, its power-law behavior and semiconductor behavior can be effectively modulated.
This enables efficient thermometry spanning from millikelvin temperatures to room temperature, and allows for precise local temperature measurement. Furthermore, this thermometer exhibits excellent temperature sensitivity and resolution, and can be fine-tuned to show small magnetoresistance.
In summary, the Ta$_{2}$Pd$_{3}$Te$_{5}$ thermometer, also referred to as a topological thermometer, exhibits outstanding performance and significant potential for measuring a wider range of temperatures compared to conventional low-temperature thermometers.

\end{abstract}

\maketitle
\noindent\textbf{1. Introduction}
\vspace{3ex}

Topological materials and physics have become a prominent field in condensed matter physics due to the emergence of various novel physical phenomena and the dissipationless transport features of topological surface/edge states \cite{TI_RMP2010,WeylDiracSM_RMP2018,MagneticTI_nature2022}.
The potential applications of these topological properties, such as topological electronic devices \cite{CriteriaAPPofTM_AM2020,Topologicalelectronics_CP2021} and topological quantum computations \cite{NonAbielianandTQC_RMP2008,TopologicalSpintronics_FP2019}, continue to be actively pursued.
However, they either require further research to fully understand their capabilities or seem to have not yet demonstrated significant technological advantages compared to current technologies.
In this context, a novel property of topological materials has been developed as a resistance thermometer for wide-range thermometry, representing a notable application for topological materials and potentially referred to as a topological thermometer.

\begin{figure*}[!thb]
\begin{center}
\includegraphics[width=7in]{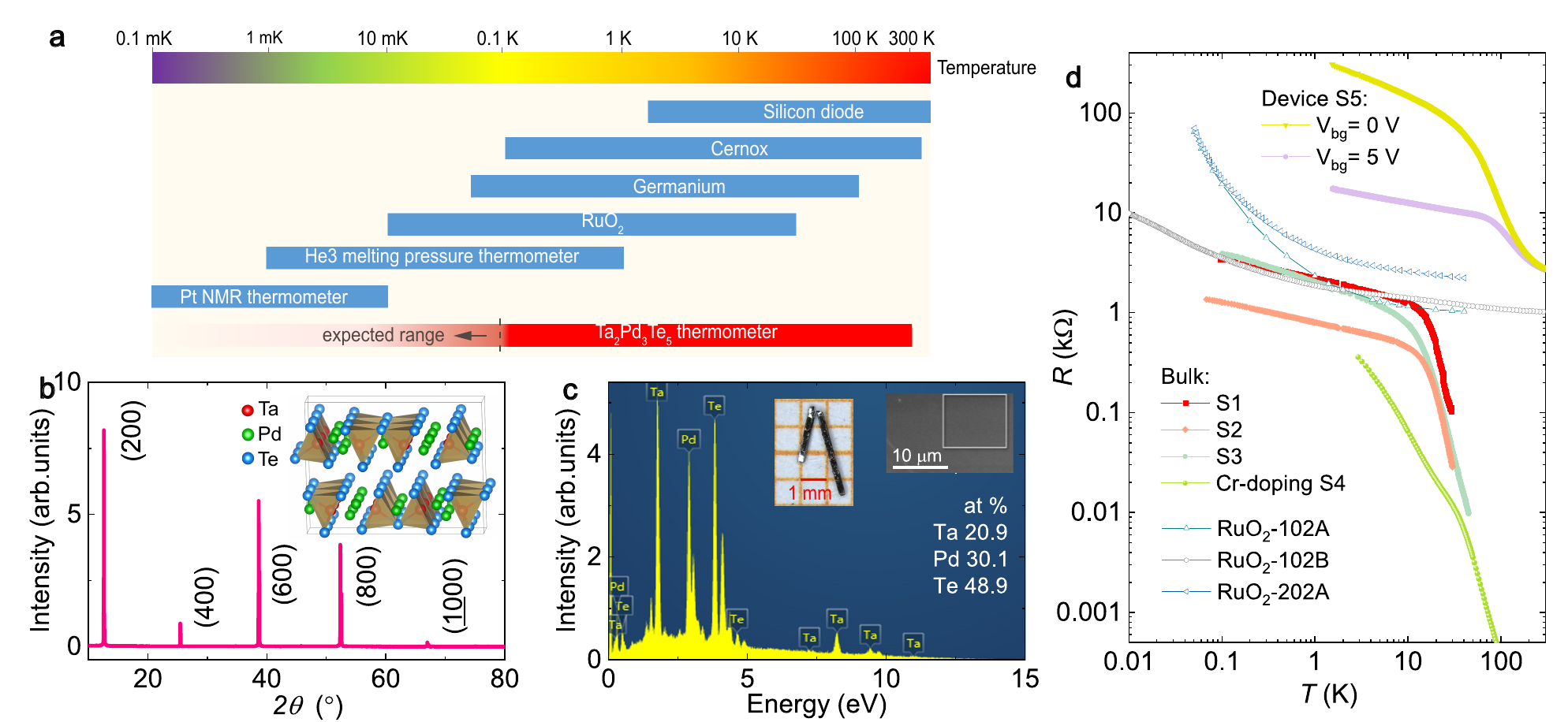}
\end{center}
\caption{\label{Fig1} Properties of Ta$_{2}$Pd$_{3}$Te$_{5}$.
a) The temperature range application for various low-temperature thermometers.
The blue bars represent commonly used commercial thermometers, while the left part of the dashed line in the Ta$_{2}$Pd$_{3}$Te$_{5}$ thermometer signifies its the predictable ability
and requires further measurements.
The temperature color bar scale is logarithmic.
b) Typical XRD spectrum for the ($l$00) facet of the single crystal.
The inset displays the structure of the vdW material with quasi-1D Ta-Te chains.
c) Typical EDS spectra, which is collected on the flat clean surface, as shown in the right inset.
The left inset displays the photomicrograph of the crystals.
d) Electrical resistance of pristine undoped and Cr-doped Ta$_{2}$Pd$_{3}$Te$_{5}$ bulk thermometers (S1 - S4),
thin-film Ta$_{2}$Pd$_{3}$Te$_{5}$ thermometer (S5), and RuO$_{2}$ thermometers.
   }
\end{figure*}

The resistance thermometer, with its ability to measure a wide temperature range from millikelvin temperatures or below to room temperature, has gained increasing significance in the field of low-temperature physics.
It plays a crucial role in efficiently studying successive phase transitions, particularly those involving ultra-low-temperature exotic quantum states such as the fractional quantum Hall effect \cite{NonAbielianandTQC_RMP2008}, anomalous quantum Hall effect \cite{QAHL_ChangCZ2013}, quantum spin liquid \cite{QSLstate_ZhouY2017}, quantum phase transitions \cite{quantumPT_RPP2003}, non-Abelian anyons \cite{NonAbelian_Nature2010}, and quantum states for quantum computing \cite{Quantumcomputing_RPP1998}, among others.
Although refrigerators supporting sub-mK temperatures or lower have been successfully developed for various types of experimental measurements \cite{500mKproperty_NC2020,100mKrefrigerator_RSI2021,NoisetThermometry_APL2013},
their widespread adoption is challenging due to the complex operations and inherent disadvantages.
One important disadvantage is that thermometers such as the He-3 melting pressure thermometer\cite{He3MeltingThermo_JLTP2002,100mKrefrigerator_RSI2021}, Coulomb blockade thermometry \cite{Coulombblockadetherm_JLTP2021,500mKproperty_NC2020}, noise thermometry \cite{NoisetThermometry_APL2013,Noisethermometry_MST2001}
and pulsed platinum NMR thermometer \cite{PtNMR_RSI1978,PtNMR_JLTP1992,NoisetThermometry_APL2013},
host complex preparation requirements and most of them are quite time-consuming for measuring ultra-low temperatures.
On the contrary, semiconductor thermometers are extensively employed at low temperatures due to their high sensitivity and active feedback temperature control, as depicted in Figure 1a.
The semiconducting feature with a negative temperature coefficient causes the resistance ($R$) to increase exponentially as the temperature ($T$) decreases.
Nevertheless, the giant resistance exhibited at millikelvin temperatures renders most resistance thermometers unusable.
Consequently, temperature detection ranges and $R$ - $T$ curves of thermometers can be modulated by various methods,
including controlling the ratio of conducting zirconium nitride to insulating zirconium oxide for Cernox thermometers \cite{Cernox_AIPConf2003}, and adjusting the RuO$_{2}$ grain size and the volume fraction of RuO$_{2}$ in the paste for RuO$_{2}$ thermometers \cite{RuO2fabrication_MST2006}.
However, the ability to detect a wide range of temperatures is often associated with a limited range of temperature measurement, as exemplified by the conventional RuO$_{2}$ thermometer typically operating within the range of 10 mK to 40 K \cite{LakeShore}.
A common approach involves using different types of thermometers to detect various temperature ranges.
Therefore, pursuing a type of resistance thermometer that deviates from the conventional semiconducting feature is crucial not only to extend the measurement limit of ultra-low temperatures but also to broaden the range of temperature measurement. This thermometer can also benefit the study of successive phase transitions or temperature-dependent physics.

In this work, a topological insulator Ta$_{2}$Pd$_{3}$Te$_{5}$ \cite{Ta2Pd3Te5TI_GuoZP_npjqm2022,Ta2Pd3Te5LL_arXiv2022,Ta2Pd3Te5QSH_GZP_PRB21} is chosen as the main material for the potential resistance thermometer.
It is a van der Waals (vdW) material with a space group $Pnma$ and features
quasi-one-dimensional (1D) tantalum-tellurium atomic chains \cite{Ta2Pd3Te5QSH_GZP_PRB21} (inset of Figure 1b).
The material exhibits several compelling properties, including excitonic insulator states \cite{Ta2Pd3Te5ExcitionIn_PRX2024,Ta2Pd3Te5ExcitionEdge_arXiv2023,Ta2Pd3Te5ExcitionIn2_PRX2024} and edge states \cite{Ta2Pd3Te5LL_arXiv2022,Ta2Pd3Te5STM_WangXG_PRB2021,Ta2Pd3Te5ExcitionEdge_arXiv2023,JDETa2Pd3Te5_arXiv2023}, which hold potential for practical applications.
For example, the edge states of the film
can be utilized in fabricating an interfering Josephson diode \cite{JDETa2Pd3Te5_arXiv2023},
which exhibits high diode efficiency at small magnetic fields with ultra-low power consumption.
This exceptional performance makes it suitable for potential applications in superconducting quantum circuits.
On the other hand, its $R$ - $T$ curve at low temperature deviates from a high-temperature semiconductor behavior ($R$ $\propto$ $e^{\bigtriangleup/2k_{B}T}$) and instead shows a power law behavior ($R$ $\propto$ $T^{-\alpha}$), where $\bigtriangleup$ is the semiconductor gap, $k_{B}$ is the Boltzmann constant, and $\alpha$ is a power law coefficient.
The power-law behavior may originate from the Luttingur liquid of edge states at low temperatures \cite{Ta2Pd3Te5LL_arXiv2022}, leading to a slower rate of resistance increase with decreasing temperature compared to the typical observation in a semiconductor.
Hence, the unique property of a Ta$_{2}$Pd$_{3}$Te$_{5}$ thermometer can be utilized for detecting lower temperatures, while retaining the traditional semiconductor behavior for high-temperature detection.
The ease of fabrication and stable performance across various thicknesses further positions the Ta$_{2}$Pd$_{3}$Te$_{5}$ thermometer as a preferred option over other Luttinger liquid systems \cite{LLcarbon_Nature1999,LuttingerLInAsGaSb_PRL2015,LLQSHE_NP2020,ChiralLLFQHE_PRL1996,ChiralLL_NP2017},
like carbon nanotubes \cite{LLcarbon_Nature1999}.

\begin{figure}[!thb]
\begin{center}
\includegraphics[width=3.5in]{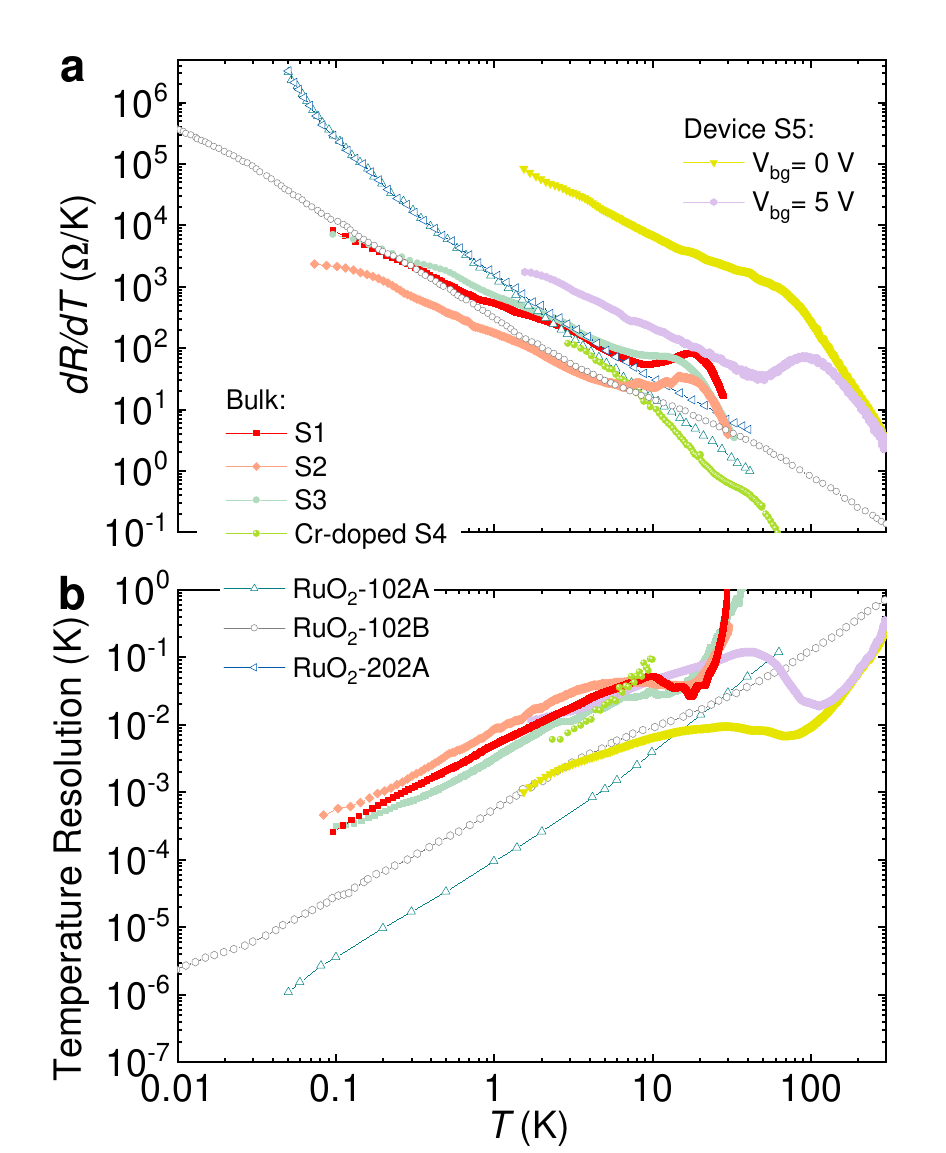}
\end{center}
\caption{\label{Fig2} a) Temperature sensitivity and b) resolution for Ta$_{2}$Pd$_{3}$Te$_{5}$ thermometers
and RuO$_{2}$ thermometers.
The data for RuO$_{2}$ thermometers originate from literatures \cite{RuO2102B_AIPCP2008,LakeShore}.
 }
\end{figure}

\vspace{3ex}
\noindent\textbf{2. Results}

\noindent\textbf{2.1. Characterization of Ta$_{2}$Pd$_{3}$Te$_{5}$}

Before introducing the Ta$_{2}$Pd$_{3}$Te$_{5}$ thermometer, we first characterize the Ta$_{2}$Pd$_{3}$Te$_{5}$ single crystal.
In Figure 1b, the typical x-ray diffraction (XRD) data exhibit
the good spectra for the ($l$00) facet of the single crystal, indicating the high quality of our samples.
Figure 1c displays a typical pattern of the energy-dispersive x-ray spectroscopy (EDS),
and the chemical composition is Ta:Pd:Te = 2.09:3.00:4.87 (Ta:Pd:Te:Cr = 1.77:3.00:4.99:0.01) for the pristine undoped (chromium-doped) compounds. Despite the low content of chromium in Cr-doped samples,
their distinct transport properties compared to pristine samples will indicate the effective Cr-doping.
Bulks and devices are synthesized or fabricated using the same method as described in the literatures \cite{Ta2Pd3Te5QSH_GZP_PRB21,Ta2Pd3Te5LL_arXiv2022}.

Transport measurement is then carried out to study the properties of Ta$_{2}$Pd$_{3}$Te$_{5}$ thermometers (Figure 1d).
The temperature-dependent resistance of bulk samples S1, S2 and S3
distinctly deviates from the semiconductor behavior below 20 K.
Notably, this deviation is integrally shown in device S5 over a wide temperature range,
leading to a power-law behavior at low-temperatures.
The power-law behavior can also be observed from $dV/dI$ - $V_{bias}$ curves and the conductance can be well scaled,
suggesting the presence of Luttinger liquid stemming from edge states \cite{Ta2Pd3Te5LL_arXiv2022}.
This type of Luttinger liquid has also been observed in other 2D systems with edge states, such as quantum spin Hall insulators \cite{LuttingerLInAsGaSb_PRL2015,LLQSHE_NP2020} and quantum Hall systems \cite{ChiralLLFQHE_PRL1996,ChiralLL_NP2017}.
The power-law behavior provides a notable advantage, as it exhibits a much smaller rate of resistance increase with decreasing temperature compared to the majority of commercial semiconductor thermometers \cite{ReviewThermometry_IEEESJ2001,LakeShore} (Figure 1d or supplementary Figure S1).
For instance, commercial thermometers like RuO$_{2}$, germanium RTD, and Cernox can easily exceed 1 M$\Omega$ at temperatures below 100 mK, rendering them unusable.
Among these conventional semiconductor thermometers, RuO$_{2}$-102B \cite{RuO2102B_AIPCP2008} exhibits the smallest resistance at ultra-low temperature (Figure 1d) and shows potential for lower temperature detection.
Compared to RuO$_{2}$-102B, the Ta$_{2}$Pd$_{3}$Te$_{5}$ thermometer possesses a much smaller increase rate of resistance at low temperature due to the small $\alpha \sim$ 0.21, 0.21 and 0.31 for samples S1, S2 and S3, respectively.
It also hosts a larger increase rate of resistance at high temperatures due to its excellent semiconductor behavior,
in comparison to most RuO$_{2}$ thermometers, suggesting its high sensitivity at high temperature.
Therefore, it is the ideal candidate for wide-range temperature detection, including ultra-low temperatures.
The small Cr-doped Ta$_{2}$Pd$_{3}$Te$_{5}$ shows low resistance at high temperatures, but it displays an approximate
power-law behavior with a large $\alpha$ value of 1.55 (see Figure 1d and supplemental Figure S2).
This behavior suggests a comparable detection limit to RuO$_{2}$-102A and RuO$_{2}$-202A, because its projected resistance approaches the M$\Omega$ range at $\sim$ 10 mK (see Table \ref{All}).

\begin{figure*}[!thb]
\begin{center}
\includegraphics[width=7in]{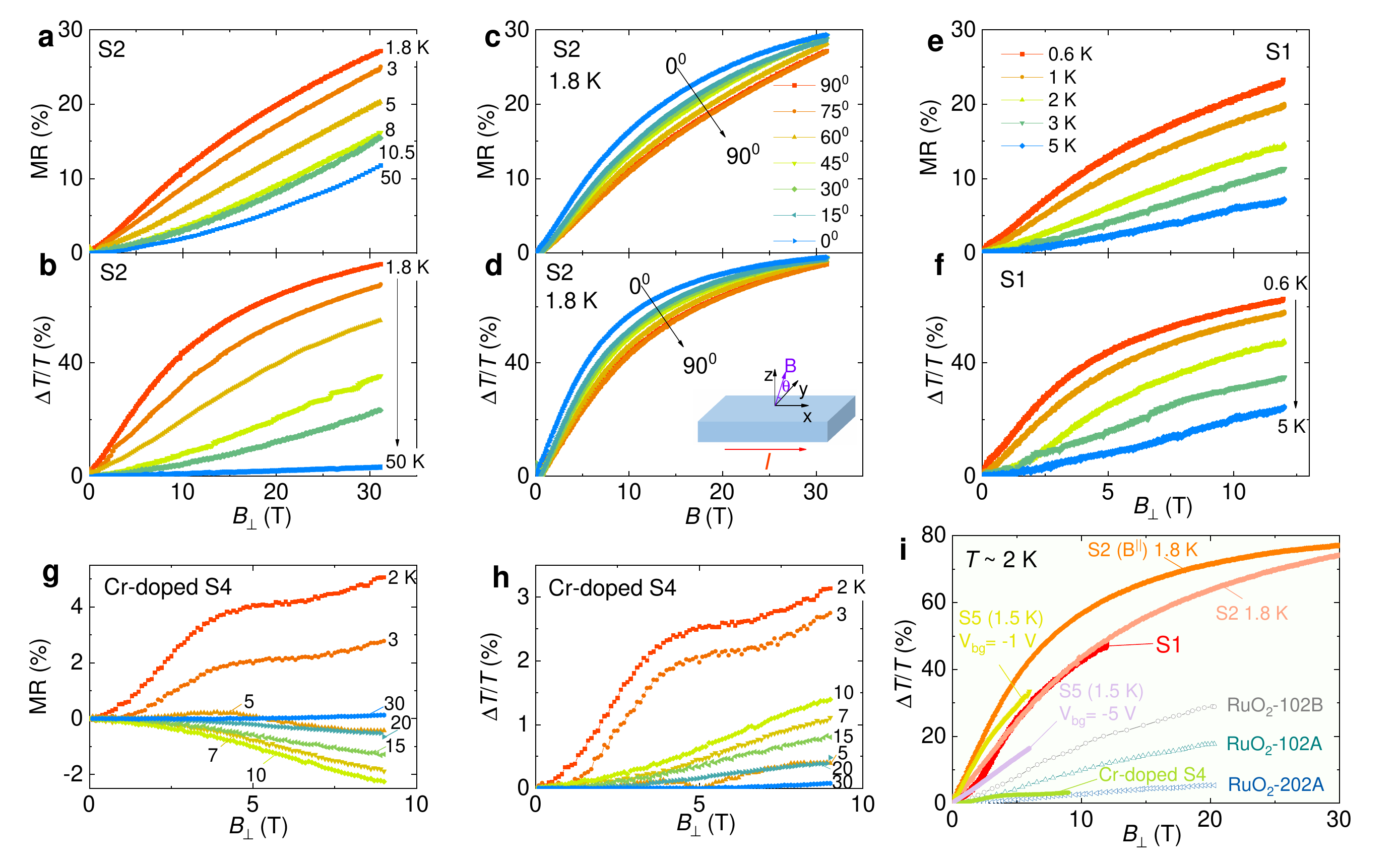}
\end{center}
\caption{\label{Fig3} Magnetoresistance of Ta$_{2}$Pd$_{3}$Te$_{5}$ thermometers.
MR of samples S2 (a), S1 (e) and Cr-doped S4 (g) at different temperatures.
MR of the Cr-doped sample S4 is much small. c) Slight anisotropic MR of sample S2.
The change of temperature [$\bigtriangleup T = |T(B) - T(0)|$] under magnetic fields for samples S2
(d), S1 (f) and S4 (h). $\bigtriangleup T/T$ below 3 K for sample S4 and S5 is approximatively estimated
based to the extended power law behavior of $R$ - $T$ curves.
i) Summary of $\bigtriangleup T/T$ ($\sim$ 2 K) under applied magnetic fields for Ta$_{2}$Pd$_{3}$Te$_{5}$
and commercial RuO$_{2}$ \cite{LakeShore} thermometers.
}
\end{figure*}

\vspace{3ex}
\noindent\textbf{2.2. Temperature sensitivity and resolution}

The temperature sensitivity of Ta$_{2}$Pd$_{3}$Te$_{5}$ thermometers is further studied in Figure 2a.
The temperature sensitivity, which describes the resistance sensitivity of the thermometer to temperature,
is typically small for semiconductor thermometers at high temperatures and significantly large at low temperatures.
For bulk Ta$_{2}$Pd$_{3}$Te$_{5}$ thermometers,
the sensitivity varies from $\sim$ 10 (20 K) to 4 $\times$ 10$^4$ $\Omega$/K (0.1 K),
and is close to that of the RuO$_{2}$-102B \cite{ReviewThermometry_IEEESJ2001}.
Their sensitivity decreases quickly above 20 K due to the gradual transition into a semiconductor state, causing a rapid reduction of resistance and resulting in weakened sensitivity.
In addition, the Cr-doped Ta$_{2}$Pd$_{3}$Te$_{5}$ thermometer also shows low sensitivity at high temperatures
and high sensitivity at low temperatures (see Figure 2a and supplemental Figure S2),
indicating its better applicability at relatively low temperatures.
Moreover, the thin-film thermometer (device S5) demonstrates higher resistance and sensitivity across a wider temperature range when compared to bulk thermometers and RuO$_{2}$ thermometers, as shown in Figure 2a and supplemental Figure S5.
These results demonstrate its suitable sensitivity for detecting a wider range of temperatures.

\begin{figure*}[!thb]
\begin{center}
\includegraphics[width=5in]{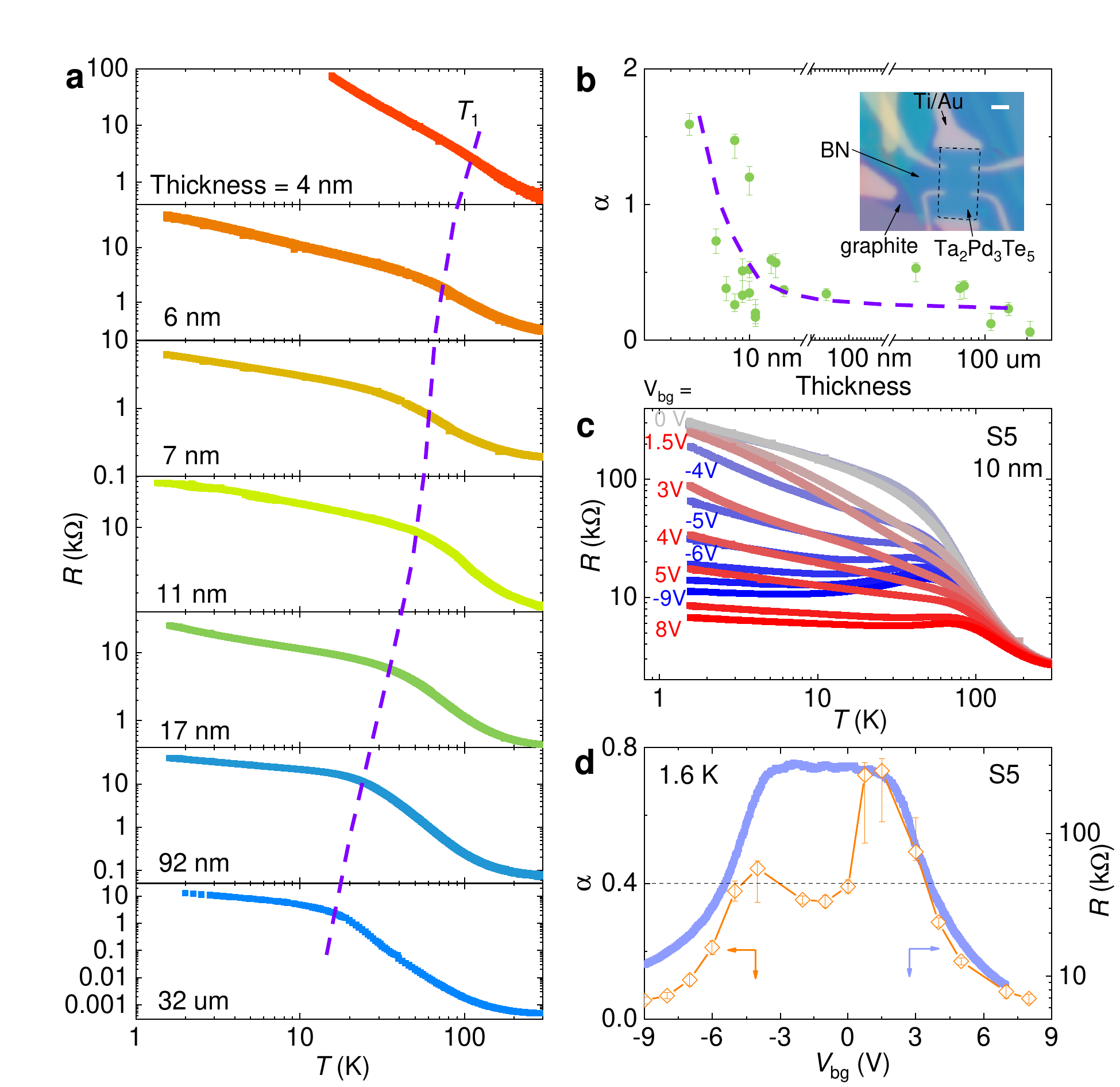}
\end{center}
\caption{\label{Fig4} Thickness and gate dependent resistance of Ta$_{2}$Pd$_{3}$Te$_{5}$.
a) Thickness dependence of $R$ - $T$ curves in Ta$_{2}$Pd$_{3}$Te$_{5}$ at CNP.
The violet dashed line is the guide line for $T_{1}$.
b) $\alpha$ as a function of thickness. The inset shows a typical Ta$_{2}$Pd$_{3}$Te$_{5}$ device,
with the white scale bar corresponding to 2 $\mu$m.
The violet dashed line serves as the guide line for thickness-dependent $\alpha$.
c) Temperature-dependent resistance at different $V_{bg}$ for device S5.
d) $V_{bg}$-dependent resistance and $\alpha$ in sample S5.
This device with $\alpha <$  0.4 seems much suitable to detect lower temperatures (Table \ref{All}).
    }
\label{Fig4}
\end{figure*}

Temperature resolution of the Ta$_{2}$Pd$_{3}$Te$_{5}$ is also shown in Figure 2b.
Temperature resolution refers to the minimum detectable change in temperature when measuring it \cite{ReviewThermometry_IEEESJ2001}.
It is influenced not only by the noise of the measuring instrument but also by the voltage or current excitation.
In the case of instrument noise, several techniques like optimized filters can be utilized to diminish system noise and improve measurement precision.
Regarding the excitation, the application of relatively large excitation in the circuit can also improve the measurement resolution.
For instance, in our systems, multilevel filters are used to minimize the system noise ($U_{n}$)
to as low as $\sim$ 10 nV.
The activation energy caused by applied current excitation ($I_{ac}$) is slightly lower than the thermal excitation energy at the given temperature,
ensuring no noticeable Joule heating effect.
Moreover, the lock-in technique (LI5650, NF Corporation, which offers the high precision measurement capabilities)
is applied to provide the excitation and measure the thermometers.
In comparison to traditional semiconductor thermometers, Ta$_{2}$Pd$_{3}$Te$_{5}$ thermometers also host relatively
high temperature resolution at low temperatures.
For bulk thermometers such as sample S1 at 0.1 K, the temperature resolution and precision (a ratio of temperature resolution to temperature) are lower than 0.3 mK and 0.3 \%, respectively.
The thin-film thermometer S5 also exhibits excellent temperature resolution at high temperatures,
highlighting its effectiveness for measuring temperatures across a wide range.
Similarly, the Cr-doped sample S4 exhibits a low-temperature resolution comparable to that of the pristine sample,
as shown in Figure 2b and supplemental Figure S2. The resolution of Cr-doped samples is only displayed at low temperatures,
as the application of a discontinuous excitation current at high temperatures leads to their discontinuous behaviors.
In summary, Ta$_{2}$Pd$_{3}$Te$_{5}$ thermometers not only demonstrate exceptional capabilities for detecting a wide range of temperatures but also exhibit excellent temperature sensitivity and resolution.

\vspace{3ex}
\noindent\textbf{2.3. Magnetoresistance}

The Ta$_{2}$Pd$_{3}$Te$_{5}$ thermometers are subsequently calibrated under a magnetic field up to 31 T.
The response of the thermometers to a magnetic field is also a crucial property,
and their small magnetoresistance is beneficial for their applications.
In Figure 3a, the magnetoresistance [MR = $R(B)$/$R(0)$ - 1] of sample S2 could reach 29\% (10\%) at 31 T (9 T) and 1.8 K, which is much smaller than that of many semimetals \cite{TransportSTM_ARMR2019,W2As3_LiYP_PRB18,NbAs2SC_Liyp,NbP_PRB_WZ}
and the germanium thermometer at low temperatures \cite{Germanium1_APL2007,LakeShore}.
The temperature display error of the thermometer due to the presence of a magnetic field can be described by a ratio
$\bigtriangleup T/T$ \cite{ReviewThermometry_IEEESJ2001} shown in Figure 3b, where $\bigtriangleup T = |T(B) - T(0)|$.
The value of $\bigtriangleup T/T$ at 1.8 K and 31 T is slight over 70\%,
indicating not a small value compared to general thermometers.
If considering a magnetic field strength of 9 T,
it is estimated that $\bigtriangleup T/T$(9 T) is 41\% at 1.8 K and 3.6\% at 10.5 K, suggesting that the magnetic effect of the Ta$_{2}$Pd$_{3}$Te$_{5}$ thermometer can be considered negligible above 10 K.

In addition, angle-dependent magnetoresistance measurements are performed to study whether there is
small magnetoresistance at specific angles.
In Figure 3c,d, MR at 1.8 K for different angle $\theta$ shows a similar situation,
with the smallest MR occurring at the perpendicular magnetic field $B_{\perp}$,
which is the most commonly used direction in this material ($B_{\perp}$ $\parallel$ $a$ axis).
MR at lower temperatures is further measured in sample S1,
leading to slight larger MR and $\bigtriangleup T/T$ in Figure 3e, f.

The magnetoresistance of Ta$_{2}$Pd$_{3}$Te$_{5}$ thermometers can be effectively reduced by adjusting the Fermi level,
for example through chemical doping or voltage gating.
The Cr-doped samples with higher carrier density exhibit a significant reduction in MR,
with $\bigtriangleup T/T$ decreasing to as low as $\sim$ 3\% at 2 K and 9 T,
as shown in Figure 3g, h and supplementary Figure S3, S4.
This observation is further supported by additional samples S6 and S7 (refer to supplemental Figure S3).
Its $\bigtriangleup T/T$ under magnetic field can be comparable to that of the RuO$_{2}$-202A \cite{LakeShore} (Figure 3i), suggesting its strong applicability under magnetic fields and indicating its potential to replace the RuO$_{2}$-202A and several resistance thermometers at ultra-low temperatures.
The reproducible measurements performed in three Cr-doped Ta$_{2}$Pd$_{3}$Te$_{5}$ thermometers highlight their high stability (see Figure 3i and supplemental Figure S3).
The thin-film thermometer also shows greater magnetoresistance than RuO$_{2}$ thermometers \cite{LakeShore}
(Figure 3i), but its magnetoresistance can be greatly reduced when
deviating from the charge neutral point (CNP), similar to the Cr-doped case.

\begin{table*}[!hbt]
\tabcolsep 0pt \caption{\label{All} Summary of ultra-low temperature thermometers.
It is important to note that values in italics in the temperature columns are the estimated resistance based on the power law or exponential behavior of the $R$ - $T$ curves, and the unit of resistance in the temperature columns is k$\Omega$. The $\bigtriangleup T/T$ columns indicate the magnetic effect of the thermometers,
and the thermometer precision is a ratio of temperature resolution to temperature.  } \vspace*{-12pt}
\begin{center}
\def\temptablewidth{2.05\columnwidth}
{\rule{\temptablewidth}{1pt}}
\begin{tabular*}{\temptablewidth}{@{\extracolsep{\fill}}cccccccc}
Sample             &$\alpha$ &$\bigtriangleup T/T$ (2 K, $B_{\bot}$ = 9 T) &Precision (0.1 K) &0.1 K &10 mK    &1 mK    &0.1 mK\\\hline
Bulk S1            &0.21     &40\%    &0.3\% &3.4    & \slshape{5.8}         &\slshape{9.5}   &\slshape{15.4}         \\
S2            &0.21     &41\% (54\%, $B^{||}$)   &0.6\% &1.3    &\slshape{2.1}         &\slshape{3.3}   &\slshape{5.4}         \\
S3            &0.31     &--    &0.3\% &3.9    &\slshape{8.4}         &\slshape{17.0}   &\slshape{34.4}         \\
Cr-doped S4        &1.55     &3.1\% &$\sim$0.3\% (2 K)   &82   &\slshape{3000}   &--  &--   \\
S5($V_{bg}$=3V)    &0.49 &10\% (1.6 K, 6 T) &4 $\times 10^{-4}$ (2 K) &\slshape{320} &\slshape{988} &$>$ \slshape{10$^{3}$} &-- \\
S5($V_{bg}$=4V)    &0.28     &-- &0.2\% (2 K)  &\slshape{69.3}    &\slshape{130.7}        &\slshape{246.1}         &\slshape{464.3}      \\
S5($V_{bg}$=5V)    &0.17     &-- &0.6\% (2 K) &\slshape{27.0} &\slshape{39.5}        &\slshape{57.8}            &\slshape{84.5}       \\
RuO$_{2}$-102B \cite{RuO2102B_AIPCP2008,LakeShore}   &0.46  &15\% &4 $\times 10^{-4}$  &3.6    &9.9     &\slshape{28}  &\slshape{83}  \\
RuO$_{2}$-102B \cite{RuO2downto5mK_Cryogenics2021,LakeShore} &0.38 &-- &--  &1.9       &4.6       &\slshape{11.3}  &\slshape{27.2}     \\
RuO$_{2}$-102A \cite{ReviewThermometry_IEEESJ2001,LakeShore} &--   &9\% &4 $\times 10^{-5}$  &193.8      & $>$ \slshape{10$^{3}$}      &--    &--     \\
RuO$_{2}$-202A \cite{ReviewThermometry_IEEESJ2001,LakeShore} &--   &9\% &-- &23.9      & $>$ \slshape{10$^{3}$}      &--    &--     \\
Ge RTD \cite{LakeShore}      &--   &60\% (8T) &2 $\times 10^{-5}$  &1.9  &$>$ \slshape{10$^{3}$}    &--   &--    \\
\end{tabular*}
{\rule{\temptablewidth}{1pt}}
\end{center}
\vspace*{-18pt}
\end{table*}

\vspace{3ex}
\noindent\textbf{2.4. Modulation of thin-film devices}

To expand the thermometry range of Ta$_{2}$Pd$_{3}$Te$_{5}$ thermometers, the thin-film thermometer is a favorable option due to its relatively high resistance at high temperatures, suitable $\alpha$ at low temperatures, excellent temperature sensitivity/resolution, and minimal magnetoresistance at the valence or conductance band.
The bulk thermometer, while performing well for ultra-low temperatures, seem not suitable for use above 30 K due to its low resistance and sensitivity, as illustrated in Figures 1, 2.
Therefore, two methods are employed to overcome this shortcoming: thickness modulation and gate modulation.
First, the thickness dependent $R - T$ curves at CNP are shown in Figure 4a.
At high temperatures, the large resistance ($>$ $\sim 100$ $\Omega$) of the sensor is observed for thickness $<$ 100 nm, ensuring relatively high sensitivity.
Thus thin-film thermometers can readily extend the working temperature range up to 300 K.
As for enhancing the measurement limit for ultra-low temperatures, there are two important factors to our knowledge.
One contributing factor is the suitable transition temperature ($T_{1}$) that deviates from the typical behavior observed in semiconductors. In Figure 4a, $T_{1}$ gradually decreases (indicated by the violet dashed line) with increasing thickness of Ta$_{2}$Pd$_{3}$Te$_{5}$ films,
which has been previously discussed in our prior literature \cite{Ta2Pd3Te5LL_arXiv2022}.
Another factor is the relatively small power-law coefficient $\alpha$.
The thin-film thermometer ($<$ $\sim$ 6 nm) appears to show not only a large resistance but also a large $\alpha$ in Figure 4b, suggesting that it may be far from possessing ideal properties for a low-temperature thermometer.
Therefore, thin-film thermometers with an optimal thickness range (10 nm $<$ thickness $<$ 100 nm) exhibit excellent and stable performance, characterized by suitable $\alpha$, resistance and $T_{1}$.
Moreover, their fabrication and application for a wide range of temperature detection can be easily achievable.

Second, gate or carrier density modulation is also an effective method for tuning the $R$ - $T$ curves of a film thermometer, particularly at low temperatures.
For an example, for sample S5, when the back gate voltage $V_{bg}$ exceeds -4 V or 5 V
(using boron nitride as a insulator medium for the gate) as shown in Figure 4c,
the resistance gradually increases or even decreases as temperature reduces,
following a departure from the semiconducting behavior.
Consequently, it begins to deviate from the favorable power-law behavior, although it may reenter an apparent power-law behavior at lower temperatures.
The relatively good power-law behavior is observed within the large gating range
(-4 V $<$ $V_{bg}$ $<$ 5 V) in Figure 4c.
In Figure 4d, $\alpha$ demonstrates similar variations to resistance in response to changes in $V_{bg}$.
The reduction of $\alpha$ at the conduction band is consistent with the results in the SiO$_{2}$ gate-tuned $\alpha$ \cite{Ta2Pd3Te5LL_arXiv2022}.
Therefore, by attaining an appropriate $R-T$ curve and $\alpha$ near the CNP,
it becomes feasible to achieve wide-range temperature detection (Table \ref{All}) and localized temperature sensing at the micron-scale.

\vspace{3ex}
\noindent\textbf{3. Discussion and conclusion}
\vspace{3ex}

The power law behavior of Ta$_{2}$Pd$_{3}$Te$_{5}$ renders the resistance thermometer exceptionally well-suited
for highly efficient measurements at ultra-low temperatures.
The potential applications of these thermometers hold tremendous promise for sub-mK temperature measurements.
However, before drawing a conclusion about this thermometer, it is essential to address several key points.
First, we will estimate the approximate temperature measurement limit for a resistance thermometer in common refrigerators.
For instance, in our dilution refrigerators, the voltage noise $U_{n}$ of the measurement circuit is close to 10 nV,
which approaches the measurement limit of general lock-in instruments, especially when considering the utilization of
multiple filters such as $\pi$ filter, Ag filter, RC filter, and others.
This system noise leads to an electron temperature of approximately 0.1 mK or 0.01 $\mu$eV,
suggesting the measurement limit of a resistance thermometer in this type of system.
If we consider the signal-to-noise ratio of the measurement,
the temperature measurement limit will be significantly greater than 0.1 mK.
Moveover, it is worth noting that the actual electron temperature of the sensor/chip in the refrigerator is usually much higher \cite{RuO2downto5mK_Cryogenics2021} due to the parasitic radio frequency heating and self heating.
The former can be optimized by multilevel filters,
while the latter can be reduced by decreasing the excitation current.
Therefore, 0.1 mK can be considered as an approximate theoretical limit for resistance thermometers in comparable refrigerators and circuits, even though only a limited number of refrigerators employing common cooling methods are capable of reaching such temperatures.

Second, this raises the question of how a resistance thermometer can effectively operate at low temperatures with relatively small resistance (less than 1 M$\Omega$).
In Table \ref{All} and supplementary Figure S1, it is evident that general commercial semiconductor thermometers,
such as Cernox, germanium RTD, RuO$_{2}$-102A and RuO$_{2}$-202A, have extensional resistances much larger than
1 M$\Omega$ below 10 mK, indicating their limited ability to effectively detect lower temperatures.
This implies that a traditional semiconductor thermometer is not suitable for detecting lower temperatures due to its exponential-type $R$ - $T$ response.
Subsequently, RuO$_{2}$-102B was developed through the control of the proportion of RuO$_{2}$, paste, and other components \cite{RuO2fabrication_MST2006}, resulting in a smaller resistance \cite{RuO2102B_AIPCP2008}.
Its resistance appears to deviate from exponential response (Figure 1d) and shows an approximate power-law behavior at low temperatures \cite{RuO2102B_AIPCP2008, RuO2downto5mK_Cryogenics2021}.
Therefore, this type of RuO$_{2}$ might extend the detecting limit to lower temperatures (Table \ref{All})
if the power law behavior is assumed to persist at lower temperatures.
However, further research is needed to check this behavior, considering it is not expected to behave like a typical Luttinger liquid.
Other notable drawbacks of this thermometer include its relatively large magnetic effect (Figure 3i) and its limited usefulness at high temperatures due to low sensitivity.
Alternatively, topological thermometers based on Ta$_{2}$Pd$_{3}$Te$_{5}$ with a suitable power law efficient
offer a promising solution to address the aforementioned concerns and perform effectively across a wide range of temperatures.

In summary, the utilization of a power-law behavior in Ta$_{2}$Pd$_{3}$Te$_{5}$ widens the potential applications of
topological materials in thermometers.
Its power law efficiency $\alpha$ can be effectively controlled by adjustments in thickness or carrier density,
allowing it to meet a wide range of requirements, including the detection of ultra-low temperatures and localized temperature variations.
For example, modulation of carrier density through gate voltage or doping can not only control the value of $\alpha$ but also reduce magnetoresistance.
In short, a film thermometer with a Fermi level that deviates slightly from the CNP exhibits exceptional performance in detecting a wide range of temperatures, from ultra-low temperatures to room temperatures,
and also shows relatively small magnetoresistance.
Therefore, this topological thermometer exhibits excellent overall performance and holds great application potential.

\vspace{3ex}

\noindent\textbf{4. Experimental Section}

Single crystals of Ta$_{2}$Pd$_{3}$Te$_{5}$ and Cr-doped Ta$_{2}$Pd$_{3}$Te$_{5}$ were synthesized using the self-flux method \cite{Ta2Pd3Te5QSH_GZP_PRB21}.
The XRD data were collected by monochromatic Cu $K_{\alpha1}$ radiation, and the composition ratio of the crystals was analyzed by EDS.
The Ta$_{2}$Pd$_{3}$Te$_{5}$ devices were fabricated using conventional nanofabrication technology \cite{Ta2Pd3Te5LL_arXiv2022}.
The electrical resistance measurements at low magnetic fields were performed in Oxford TeslatronPT cryostats and dilution refrigerators. The  High field magnetotransport measurements were also performed using a standard AC lock-in technique with a DC-resistive magnet ($\sim$ 31 T) at the China High Magnetic Field Laboratory (CHMFL) in Hefei.

\vspace{3ex}

\noindent\textbf{Supporting Information}

\noindent Supporting Information is available from the Online Library or from the author.

\vspace{3ex}

\noindent\textbf{Acknowledgments}

\noindent This work was supported by the Beijing Natural Science Foundation (Grant No. JQ23022),
the National Natural Science Foundation of China
(Grant Nos. 92065203, 12174430, U2032204, 11974395, 12188101 and 12122411),
the Strategic Priority Research Program of Chinese Academy of Sciences (Grant Nos. XDB33000000 and XDB33030000),
the Beijing Nova Program (Grant No. Z211100002121144),
the National Key R$\&$D Program of the MOST of China  (Grant No. 2022YFA1602602),
the Informatization Plan of Chinese Academy of Sciences (CAS-WX2021SF-0102),
the China Postdoctoral Science Foundation (Grant Nos. 2021M703462 and 2021TQ0356) 
and the Center for Materials Genome.
Additionally, a portion of this work was carried out at the Synergetic Extreme Condition User Facility (SECUF)
and China High Magnetic Field Laboratory (CHMFL) in Hefei.

\vspace{3ex}

\noindent\textbf{Competing Interests}

\noindent The authors declare no competing interests.

\vspace{3ex}

\noindent\textbf{Author Contributions}

\noindent J. S. and Y.P.L. conceived and designed the experiment.
Y.P.L. and A.Q.W. fabricated the devices with assistance of G.Y., X.C.G. and Y. H..
Y.P.L. and A.Q.W. performed the transport measurements,
supervised by T.Q., G.T.L, F.M.Q., L.L. and J. S..
J.L.Z. performed the high field transport measurements.
D.Y.Y. and Y.G.S. prepared Ta$_{2}$Pd$_{3}$Te$_{5}$ crystals.
Z.J.W. provided the theoretical support.
Y.P.L. and J.S. wrote the manuscript with helps from all other co-authors.

\vspace{3ex}

\noindent\textbf{Data Availability Statement}

\noindent The data that support the findings of this study are available
from the corresponding author upon reasonable request.

\vspace{3ex}

\noindent\textbf{Keywords}

\noindent Ta$_{2}$Pd$_{3}$Te$_{5}$, topological insulator, topological thermometer,
edge state, ultra-low temperature


\end{document}


\title{Supplementary information for \\
 Ta$_{2}$Pd$_{3}$Te$_{5}$ topological thermometer }

\author{Yupeng Li}
      \thanks{Equal contributions}
      \affiliation{Beijing National Laboratory for Condensed Matter Physics, Institute of Physics, Chinese Academy of Sciences, Beijing 100190, China}

\author{Anqi Wang}
      \thanks{Equal contributions}
      \affiliation{Beijing National Laboratory for Condensed Matter Physics, Institute of Physics, Chinese Academy of Sciences, Beijing 100190, China}
      \affiliation{School of Physical Sciences, University of Chinese Academy of Sciences, Beijing 100049, China}

\author{Dayu Yan}
      \affiliation{Beijing National Laboratory for Condensed Matter Physics, Institute of Physics, Chinese Academy of Sciences, Beijing 100190, China}

\author{Guang Yang}
      \affiliation{Beijing National Laboratory for Condensed Matter Physics, Institute of Physics, Chinese Academy of Sciences, Beijing 100190, China}

\author{Xingchen Guo}
      \affiliation{Beijing National Laboratory for Condensed Matter Physics, Institute of Physics, Chinese Academy of Sciences, Beijing 100190, China}
      \affiliation{School of Physical Sciences, University of Chinese Academy of Sciences, Beijing 100049, China}

\author{Yu Hong}
      \affiliation{Beijing National Laboratory for Condensed Matter Physics, Institute of Physics, Chinese Academy of Sciences, Beijing 100190, China}
      \affiliation{School of Physical Sciences, University of Chinese Academy of Sciences, Beijing 100049, China}

\author{Guangtong Liu}
      \affiliation{Beijing National Laboratory for Condensed Matter Physics, Institute of Physics, Chinese Academy of Sciences, Beijing 100190, China}
      \affiliation{Songshan Lake Materials Laboratory, Dongguan 523808, China}

\author{Fanming Qu}
      \affiliation{Beijing National Laboratory for Condensed Matter Physics, Institute of Physics, Chinese Academy of Sciences, Beijing 100190, China}
      \affiliation{School of Physical Sciences, University of Chinese Academy of Sciences, Beijing 100049, China}
      \affiliation{Songshan Lake Materials Laboratory, Dongguan 523808, China}

\author{Zhijun Wang}
      \affiliation{Beijing National Laboratory for Condensed Matter Physics, Institute of Physics, Chinese Academy of Sciences, Beijing 100190, China}
      \affiliation{School of Physical Sciences, University of Chinese Academy of Sciences, Beijing 100049, China}

\author{Tian Qian}
      \affiliation{Beijing National Laboratory for Condensed Matter Physics, Institute of Physics, Chinese Academy of Sciences, Beijing 100190, China}
      \affiliation{Songshan Lake Materials Laboratory, Dongguan 523808, China}

\author{Jinglei Zhang}
      \email{zhangjinglei@hmfl.ac.cn}
      \affiliation{Anhui Province Key Laboratory of Condensed Matter Physics at Extreme Conditions,
      High Magnetic Field Laboratory of the Chinese Academy of Sciences, Hefei 230031, Anhui, China}

\author{Youguo Shi}
      \email{ygshi@iphy.ac.cn}
      \affiliation{Beijing National Laboratory for Condensed Matter Physics, Institute of Physics, Chinese Academy of Sciences, Beijing 100190, China}
      \affiliation{Songshan Lake Materials Laboratory, Dongguan 523808, China}

\author{Li Lu}
      \email{lilu@iphy.ac.cn}
      \affiliation{Beijing National Laboratory for Condensed Matter Physics, Institute of Physics, Chinese Academy of Sciences, Beijing 100190, China}
      \affiliation{School of Physical Sciences, University of Chinese Academy of Sciences, Beijing 100049, China}
      \affiliation{Songshan Lake Materials Laboratory, Dongguan 523808, China}

\author{Jie Shen}
      \email{shenjie@iphy.ac.cn}
      \affiliation{Beijing National Laboratory for Condensed Matter Physics, Institute of Physics, Chinese Academy of Sciences, Beijing 100190, China}
      \affiliation{Songshan Lake Materials Laboratory, Dongguan 523808, China}


\maketitle

\begin{figure*}[!thb]
\begin{center}
\includegraphics[width=5in]{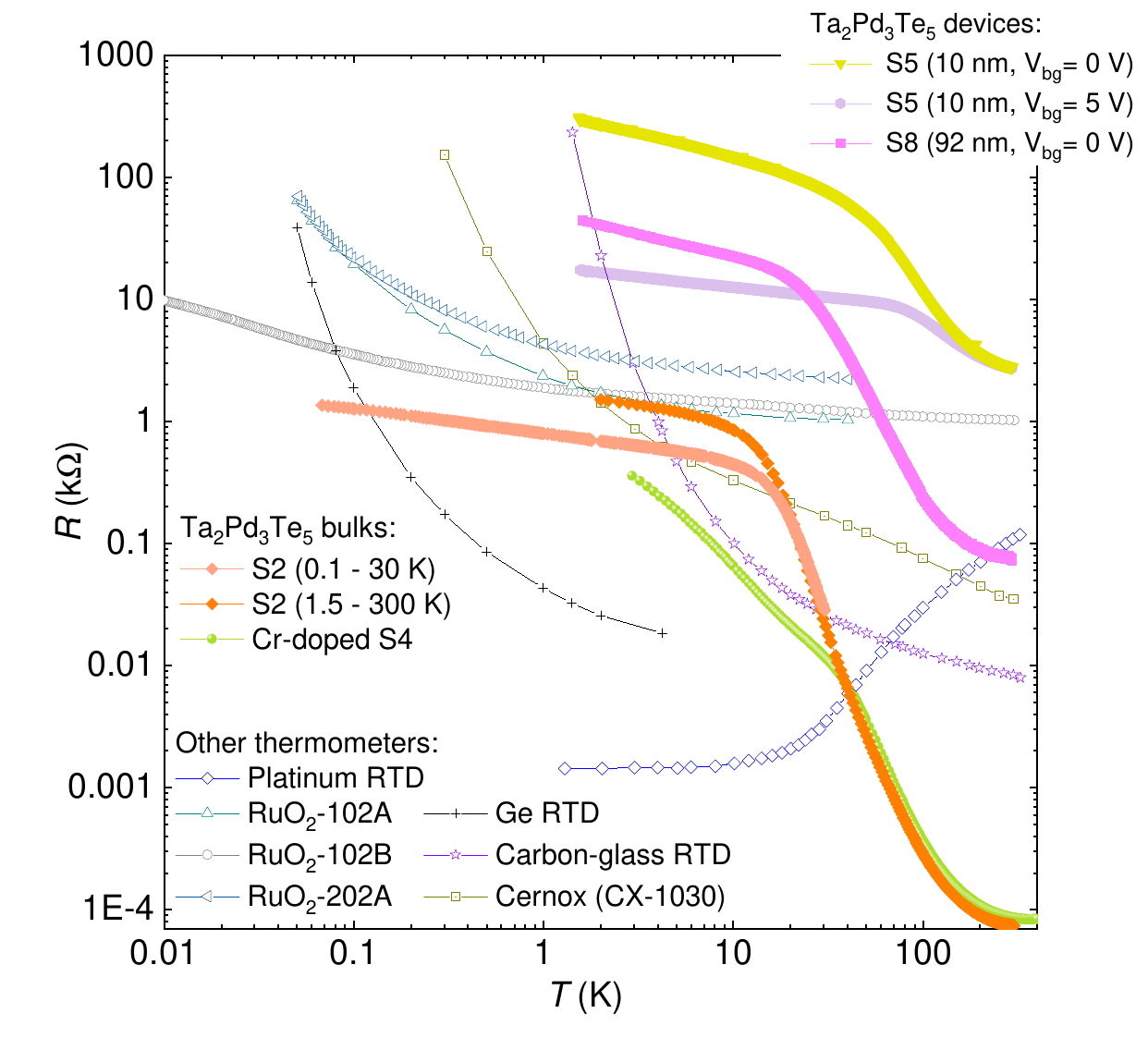}
\end{center}
\caption{\label{Fig4.5} \textbf{Summary of Ta$_{2}$Pd$_{3}$Te$_{5}$ and other thermometers.}
Temperature-dependent resistance in Ta$_{2}$Pd$_{3}$Te$_{5}$ thermometers, RuO$_{2}$ and other thermometers (open scatters) \cite{ReviewThermometry_IEEESJ2001,LakeShore} is shown.
Significantly, the rate of increase in resistance with decreasing temperature in Ta$_{2}$Pd$_{3}$Te$_{5}$ thermometers is much lower than that in conventional semiconductor thermometers, indicating their great potential for lower temperature detection. The bulk Ta$_{2}$Pd$_{3}$Te$_{5}$ sample S2 is measured at high or low temperatures under different electrode preparation conditions.
   }
\end{figure*}

\begin{figure*}[!thb]
\begin{center}
\includegraphics[width=3.5in]{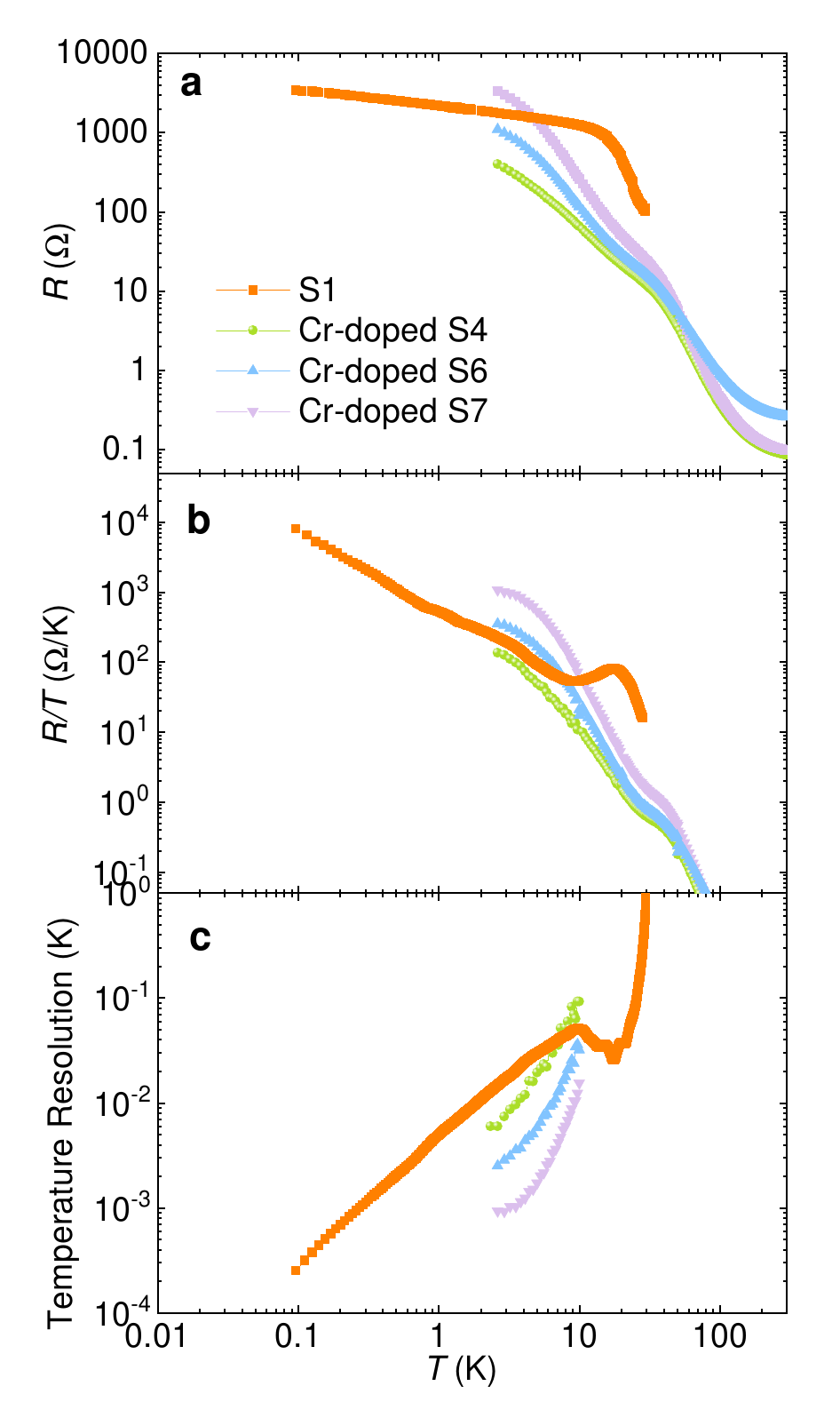}
\end{center}
\caption{\label{Fig4.5} \textbf{Properties of Cr-doped Ta$_{2}$Pd$_{3}$Te$_{5}$.}
$R$ - $T$ curves ($\mathbf{a}$), temperature sensitivity ($\mathbf{b}$) and resolution ($\mathbf{c}$) of three Cr-doped Ta$_{2}$Pd$_{3}$Te$_{5}$ samples and an undoped sample S2 are shown. The approximate power law behavior at low temperatures displays $\alpha$ = 1.93 (2.35) with an error bar of $\pm 0.05$ for sample S6 (S7). The chemical composition of the three Cr-doped Ta$_{2}$Pd$_{3}$Te$_{5}$ samples is also similar. The only low-temperature resolution of Cr-doped samples are shown, because discontinuous excitation current is applied at high temperatures, resulting in the discrete behavior. }
\end{figure*}

\begin{figure*}[!thb]
\begin{center}
\includegraphics[width=6in]{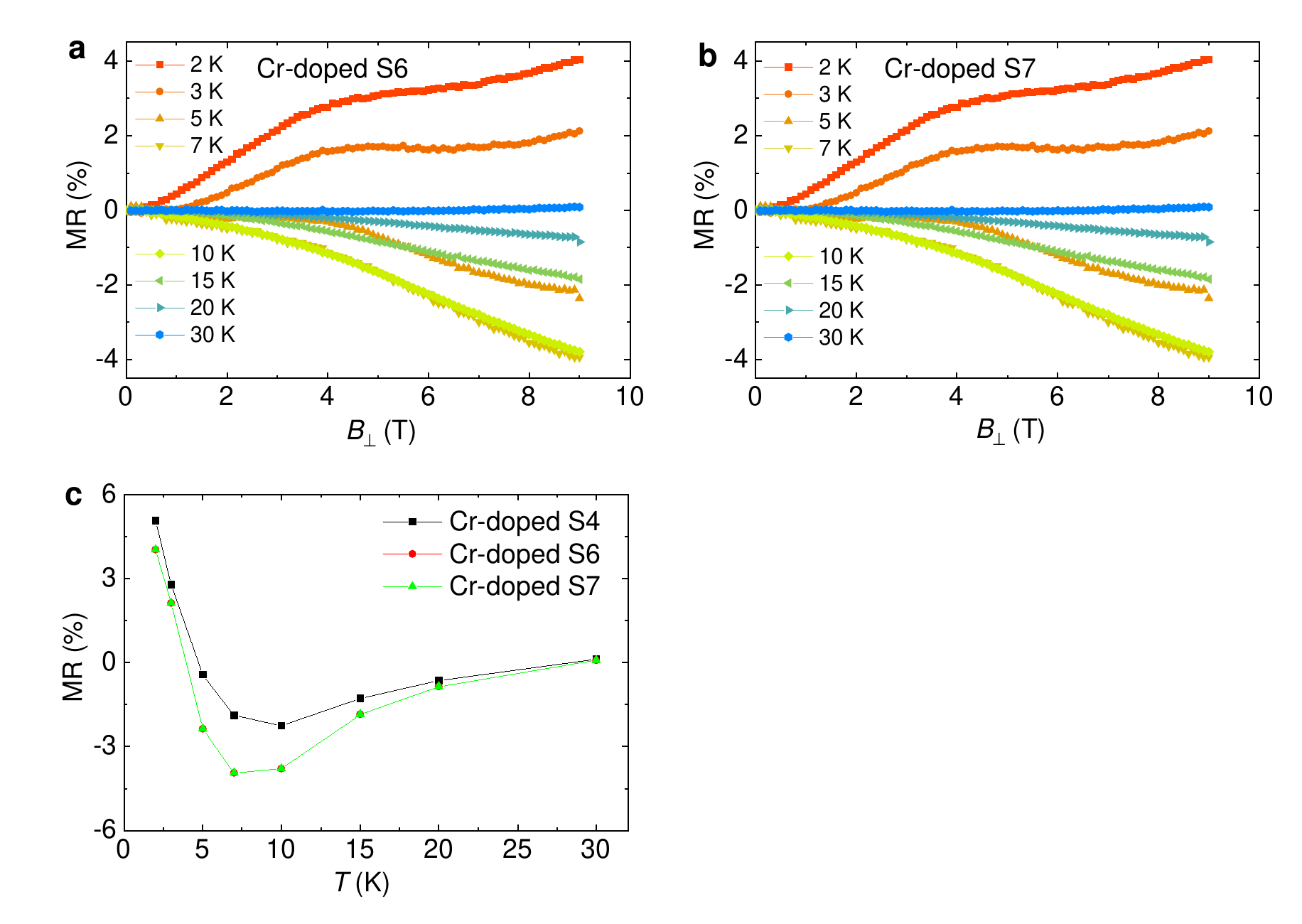}
\end{center}
\caption{\label{Fig4.5} \textbf{Magnetoresistance of Cr-doped Ta$_{2}$Pd$_{3}$Te$_{5}$.}
Magnetoresistance of Cr-doped Ta$_{2}$Pd$_{3}$Te$_{5}$ samples S6 ($\mathbf{a}$) and S7 ($\mathbf{b}$) at different temperatures. $\mathbf{c}$, Magnetoresistance of three Cr-doped samples at 9 T.
}
\end{figure*}

\begin{figure*}[!thb]
\begin{center}
\includegraphics[width=6.5in]{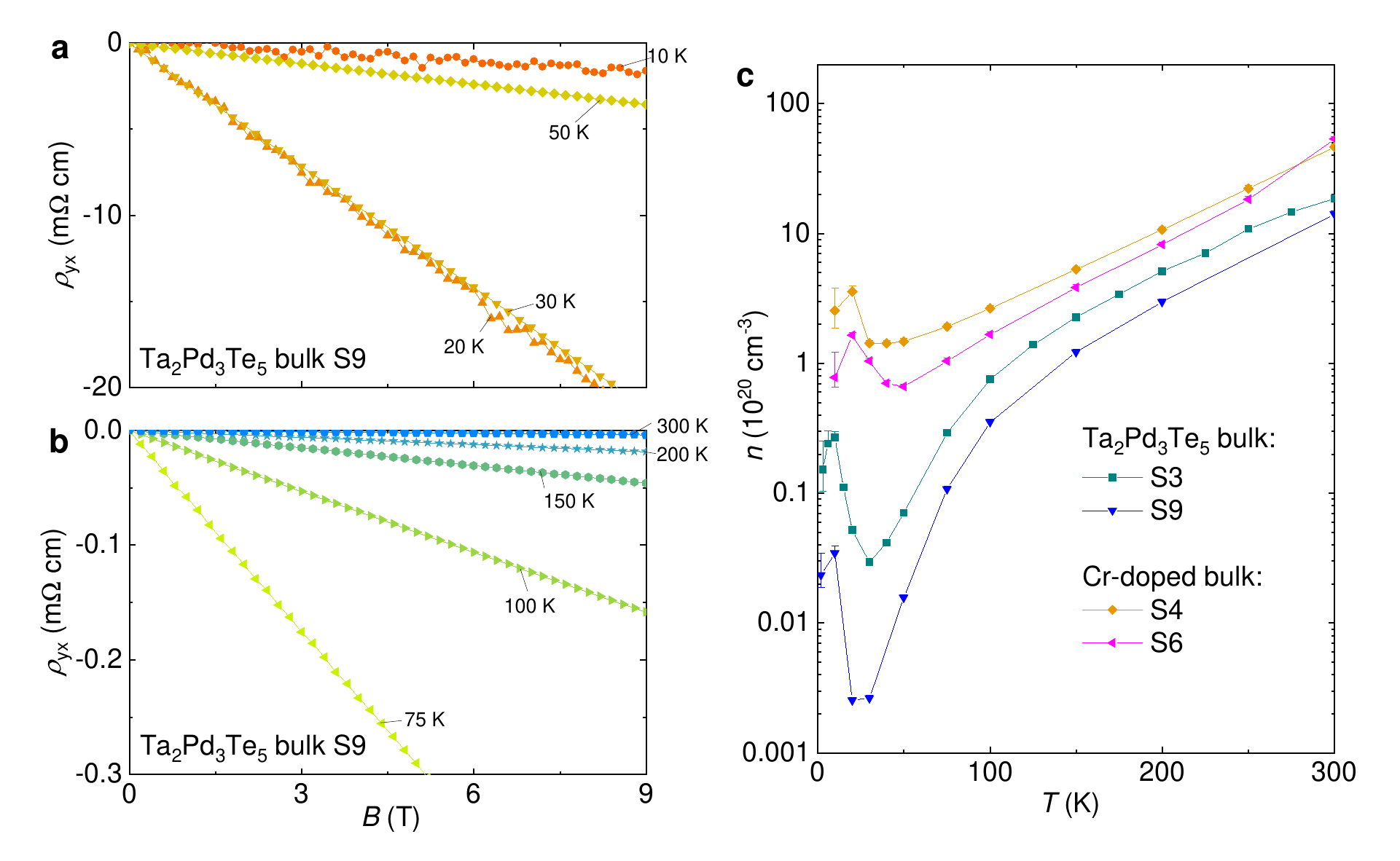}
\end{center}
\caption{\label{Fig6} \textbf{Carrier density.}
$\mathbf{a, b}$, Typical Hall resistivity of Ta$_{2}$Pd$_{3}$Te$_{5}$.
The sample S9 shows the typical election-type Hall resistivity.
$\mathbf{c}$, Temperature-dependent carrier density for pristine undoped and Cr-doped Ta$_{2}$Pd$_{3}$Te$_{5}$.
Cr-doped Ta$_{2}$Pd$_{3}$Te$_{5}$ samples display the higher carrier density than Ta$_{2}$Pd$_{3}$Te$_{5}$.
}
\end{figure*}

\begin{figure*}[!thb]
\begin{center}
\includegraphics[width=6in]{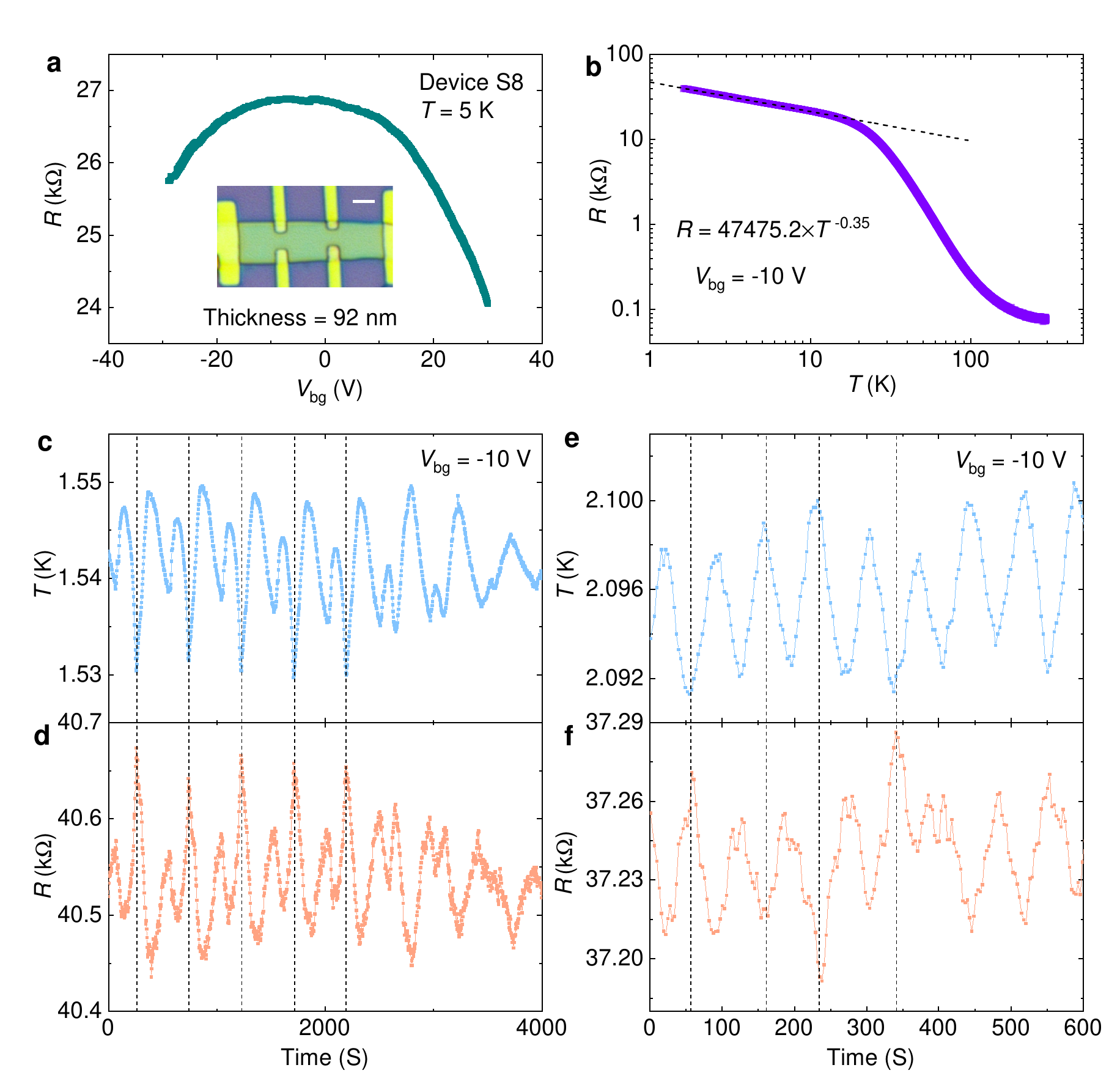}
\end{center}
\caption{\label{Fig4.5} \textbf{Temperature sensitivity of Ta$_{2}$Pd$_{3}$Te$_{5}$ device S8 as a function of time.}
$\mathbf{a}$, $V_{bg}$-dependent resistance of device S8 at 5 K.
The inset displays the photomicrograph of device S8 with a thickness of 92 nm and SiO$_{2}$/Si+ as bottom gate,
and the white scale bar corresponds to 2 $\mu$m.
$\mathbf{b}$, $R$ - $T$ curve at $V_{bg} = -10$ V with $\alpha = -0.35$.
$\mathbf{c, e}$, System temperature as a function of time at $V_{bg} = -10$ V.
$\mathbf{d, f}$, Corresponding resistance as a function of time and temperature.
The oscillation behavior of temperature and resistance is consistent (see the typical dashed black lines),
indicating high temperature sensitivity of Ta$_{2}$Pd$_{3}$Te$_{5}$ thermometers at low temperatures.
}
\end{figure*}

\clearpage